\documentclass[runningheads]{llncs}
\usepackage[T1]{fontenc}
\usepackage{graphicx}
\usepackage{amsmath}
\usepackage{multirow} 

\usepackage{soul}
\begin{document}
\title{LSTM Autoencoder-based Deep Neural Networks for Barley Genotype-to-Phenotype Prediction}
%
%

\author{Guanjin Wang\inst{1} \and
Junyu Xuan\inst{2} \and
Penghao Wang\inst{3} \and
Chengdao Li \inst{4,5,6} \and
Jie Lu \inst{2}}

\authorrunning{G.Wang et al.}

\institute{School of Information Technology, Murdoch University, Australia\and
Australia Artificial Intelligence Institute, Faculty of Engineering and Information Technology, University of Technology Sydney, Australia\and
School of Medical, Molecular and Forensic Sciences, Murdoch University, Australia\and
Western Crop Genetics Alliance, Murdoch University, Australia\and
Western Australian State Agricultural Biotechnology Centre, Australia\and
Department of Primary Industries and Regional Development, Perth, Australia}

%
\maketitle              

\begin{abstract}

Artificial Intelligence (AI) has emerged as a key driver of precision agriculture, facilitating enhanced crop productivity, optimized resource use, farm sustainability, and informed decision-making. Also, the expansion of genome sequencing technology has greatly increased crop genomic resources, deepening our understanding of genetic variation and enhancing desirable crop traits to optimize performance in various environments. There is increasing interest in using machine learning (ML) and deep learning (DL) algorithms for genotype-to-phenotype prediction due to their excellence in capturing complex interactions within large, high-dimensional datasets.
In this work, we propose a new LSTM autoencoder-based model for barley genotype-to-phenotype prediction, specifically for flowering time and grain yield estimation, which could potentially help optimize yields and management practices. Our model outperformed the other baseline methods, demonstrating its potential in handling complex high-dimensional agricultural datasets and enhancing crop phenotype prediction performance.

\end{abstract}
\section{Introduction}
AI, particularly ML and DL, has been widely applied across various industries in recent years \cite{eli2019applications}. In particular, it has become a key enabler of precision agriculture, also known as smart farming, which has transformed modern farming practices \cite{sharma2020machine,bhat2021big,adewusi2024ai}. As the global population grows and climate change intensifies, the need for sustainable farming practices has become increasingly critical \cite{mana2024sustainable}. AI techniques have demonstrated significant potential in enhancing agricultural productivity and efficiency, optimizing resource use, and making farming operations more sustainable and profitable. These techniques also enable farmers and other stakeholders to make more informed decisions \cite{mana2024sustainable,pandey2024towards}.
One of the significant applications of precision agriculture is predicting crop phenotypes from genotypes. Thanks to the expansion of genome sequencing technology, crop genomic resources have greatly increased, deepening our understanding of genetic variation and enhancing desirable plant traits to optimize performance in various environments \cite{danilevicz2022plant}.

In this work, we focus on barley (\textit{Hordeum vulgare} L.), a crucial cereal crop both globally and nationally. Barley is cultivated in highly productive agricultural areas as well as in marginal environments subject to adverse conditions \cite{baum2007molecular}. Known for its resilience compared to other cereals like wheat and rice, barley can adapt to various biotic and abiotic stresses, making it essential for maintaining and increasing production in marginal areas to ensure food security \cite{tester2010breeding}. In Western Australia, for example, barley ranks as the second-largest cereal crop, contributing 25\% of the state's total grain production and generating over \$1 billion annually in export earnings from barley grain and malt. About 30\% of this barley is classified as malting grade for the international beer industry, while the remaining 70\% is feed grade, primarily exported to the Middle East \cite{WA_Barley}.

To maximize yield and minimize exposure to environmental stresses such as frost, heat, and drought during the growing season, it is crucial for barley to flower within a specific time window \cite{maurer2015modelling}. Also genes that control flowering time, often overlap with those related to grain yield \cite{hill2019hybridisation}. Understanding the genetic data and their association with flowering time prediction and grain yield is vital for advancing barley improvement to meet future food and feed demands, enhance crop quality, and optimize management practices, including pest and disease control and harvesting schedules.
Many previous studies in this area have utilized traditional statistical methods, but recent years have seen a growing interest in ML and DL algorithms for genotype-to-phenotype prediction due to their advanced learning capabilities. These algorithms excel at capturing complex, higher-order interactions and achieving higher predictability with high-dimensional datasets, making them highly effective at linking plant genotypes with phenotypes \cite{danilevicz2022plant,lecun2015deep}. Some existing studies has demonstrated the success of ML and DL models such as ensemble learning methods, kernel-based methods, and deep neural networks in predicting a wide range of agronomic traits by capturing the intricate interactions between genotype, phenotype, and the environment \cite{danilevicz2022plant,crossa2019deep,grinberg2020evaluation,khaki2020cnn}, showcasing their significant potential.
In this work, we propose a new Long Short-Term Memory (LSTM) autoencoder-based deep nueral network model for crop genotype-to-phenotype prediction to enhance predictive performance on the complex, high-dimensional datasets, with a specific application in predicting the barley flowering time and grain yield. We use a real barley dataset that includes multi-environment field trials conducted over five diverse geographical locations across two years in Western Australia, encompassing high-dimensional genomic, phenotypic, and environmental information.

The remainder of this paper is organized as follows: Section 2 provides a brief review of crop genotype-to-phenotype prediction using AI and relevant techniques. Section 3 details the methodologies of our proposed model. Section 4 discusses the adopted dataset, experimental setup, and results. Finally, Section 5 concludes the paper.

\section{Previous Work}
This section first briefly introduces different crop genotype-to-phenotype modelling methods. Then, it reviews the LSTM model, which is used as a main component of our proposed model.

\subsection{Crop genotype-to-phenotype prediction}
Understanding plant genotype-to-phenotype relationships is crucial for improving crop performance and resilience, food security and sustainability. Linear modeling approaches such as Genomic Best Linear Unbiased Prediction (GBLUP) \cite{clark2013genomic} and Bayesian systems \cite{tong2021machine} have traditionally been used in genomic selection and genotype-to-phenotype prediction. However, these methods may face performance limitations due to the high dimensionality of marker data and the complex patterns within. 
There is growing interest in utilizing ML and DL techniques to predict plant phenotypes, as these methods can capture nuanced relationships among variables and efficiently handle large datasets, leading to improved predictive accuracy.
For instance, Ma et al. \cite{ma2018deep} demonstrated the effectiveness of Convolutional Neural Networks (CNNs) in extracting informative genomic features, thereby improving selection accuracy in plant breeding programs. 
Kick et al. \cite{kick2023yield} examined optimized Deep Neural Network (DNN) models, which produced more consistent maize yield estimates despite having a slightly higher average error than the best BLUP model. These results show the DNN's promise for complementing existing models in crop selection and improvement.
Wu et al. \cite{wu2024transformer} investigated the use of Transformer-based DNNs for genomic prediction, introducing a new model named GPformer. GPformer integrates information from all relevant SNPs, irrespective of their physical distance, to achieve a holistic understanding. Extensive experiments across five diverse crop datasets demonstrated that GPformer consistently outperformed traditional methods such as ridge regression-based linear unbiased prediction (RR-BLUP), support vector regression (SVR), light gradient boosting machine (LightGBM), and deep neural network genomic prediction (DNNGP) in terms of reducing mean absolute error. 
Kkut et al. \cite{okut2021deep} reviewed major DL approaches, including fully connected DNNs, Recurrent Neural Networks (RNNs), CNNs, and Long Short-Term Memory (LSTM) networks, as well as various variations of these architectures for complex trait genomic prediction. However, the use of DL architectures such as RNNs still remains largely unexplored in genotype-to-phenotype predictions, despite presenting a potential alternative to traditional statistical methods \cite{danilevicz2022plant}.

\subsection{LSTM}


The LSTM network, introduced by Hochreiter and Schmidhuber \cite{hochreiter1997long}, is a type of RNN model \cite{grossberg2013recurrent} that specifically addresses the vanishing gradient problem found in traditional RNNs. LSTM networks are designed with special units called memory cells that can maintain information from multiple previous layers and pass them through the network as needed, allowing them to effectively find and utilize relationships and patterns within the data.
Unlike standard RNNs, LSTM networks have a hidden layer with additional units to manage the flow of information to and from the memory cells. These units include the input unit, which determines what information should be added to the memory cell based on its high activation levels; the forget unit, which clears the memory cell when its activation is high, effectively 'forgetting' unnecessary information; and the output unit, which transfers information from the memory cell to the next neuron if it has high activation.
The mathematical functions governing these units' operations are formulated as follows \cite{yu2019review}:
\begin{align}
    f_t &= \sigma(W_{xf} \cdot x_t + W_{hf} \cdot h_{t-1} + b_f) \label{3.1}\\
    i_t &= \sigma(W_{xi} \cdot x_t + W_{hi} \cdot h_{t-1} + b_i) \label{3.2}\\
    o_t &= \sigma(W_{xo} \cdot x_t + W_{ho} \cdot h_{t-1} + b_o) \label{3.3}\\
    c'_t &= \tanh(W_{xg} \cdot x_t + W_{hg} \cdot h_{t-1} + b_g) \label{3.4}\\
    c_t &= f_t \odot c_{t-1} + i_t \odot c'_t \label{3.5}\\
    h_t &= o_t \odot \tanh(c_t) \label{3.6}
\end{align}
where \( x_t \) represents the input at time step \( t \), \( W \) are the weight parameter matrices, and \( b \) are the bias vectors. Eq. (\ref{3.5}) denotes \( c_t \), the cell state at time step \( t \), while Eq. (\ref{3.6}) denotes \( h_t \), the hidden state at time step \( t \). Here, \( \cdot \) indicates standard matrix multiplication, \( \odot \) represents the elementwise product, and \( \sigma \) is the Sigmoid function. The weights and biases remain consistent across all time steps. Eqs. (\ref{3.1})-(\ref{3.3}) describe the three gates: the input gate \( i \), forget gate \( f \), and output gate \( o \). These gates control the flow of information within the cell by generating values between 0 and 1 to write to the internal memory \( c_t \), reset the memory, or read from the memory, respectively.


\section{Methodology}

\begin{figure}[h]
\centering
\includegraphics[scale=0.36]{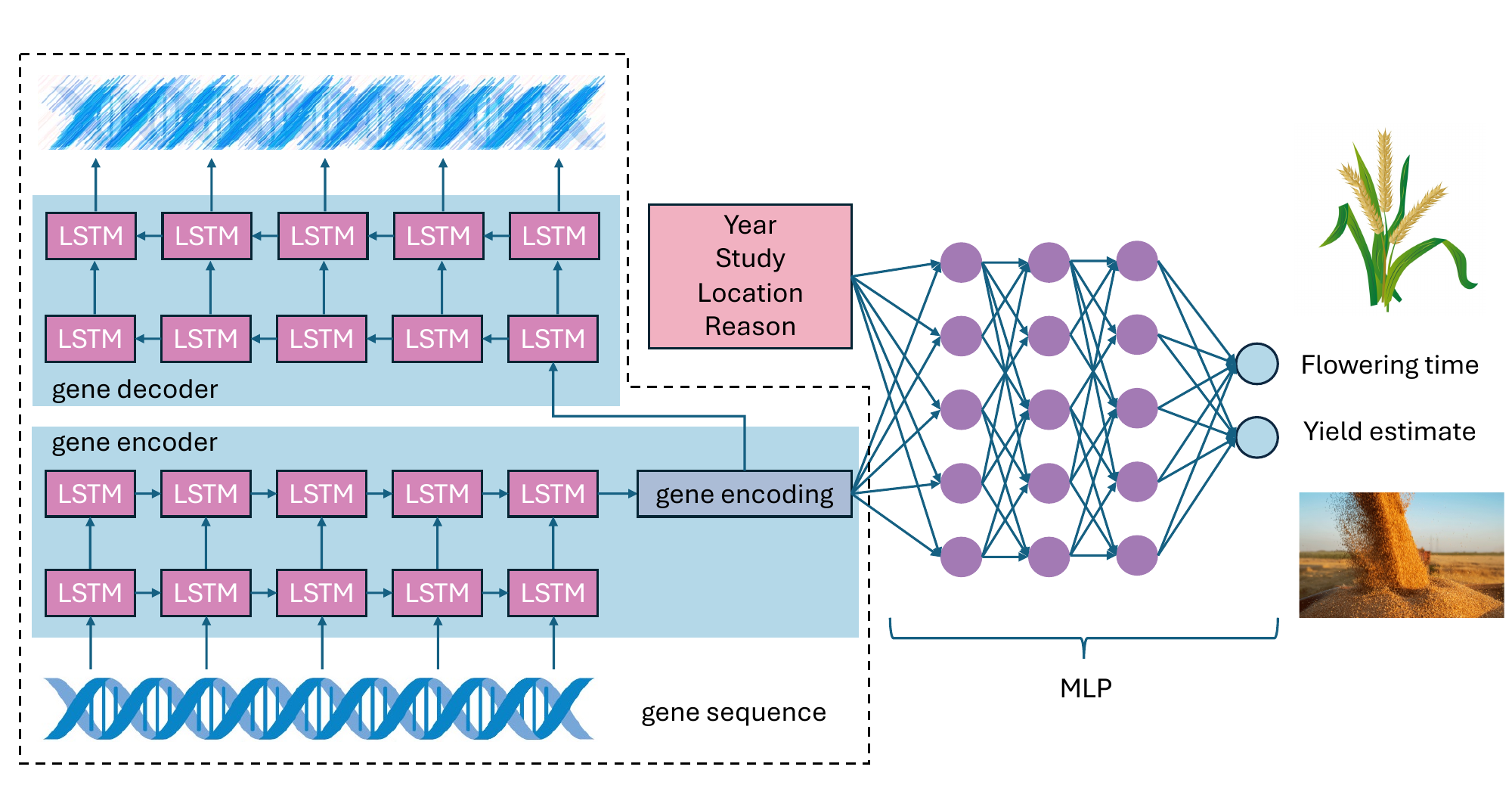}
\caption{Our LSTM autoencoder-based deep neural network framework}
\label{fig:model}
\end{figure}

In this section, we propose a LSTM autoencoder-based DNN architecture for genotype-to-phenotype prediction, as illustrated in Fig. \ref{fig:model}. It is composed of two main components: one for genomic data encoding and the other for genotype-to-phenotype prediction. Next, we are going to explain these two components in detail. 

\subsection{Genomic data encoding}
Crop genomic data encodes critical traits such as disease resistance, drought tolerance, and yield potential, making it essential for genotype-to-phenotype prediction. Our prior experiments indicated that naively using genomic data directly as features for ML models is insufficient.
In this work, we propose using an LSTM as the encoder to learn the hidden representation of high-dimensional, large-sized genomic data for each crop variant, thereby reducing the reliance on traditional feature engineering and processing.
We follow the classical setting\footnote{https://github.com/fabiozappo/LSTM-Autoencoder-Time-Series} to stack two layers of LSTM rather than one to obtain enhanced information abstraction and increase the capability to capture more complex hidden feature representation information. 

To further improve the performance, we propose to pretrain the LSTM using an autoencoder structure \cite{srivastava2015unsupervised} before the phenotype prediction. In particular, a corresponding LSTM decoder is constructed to decode the gene from its latent representation, as shown in Fig. \ref{fig:model}, using the following loss function
\begin{equation}
    \mathcal{L} = \ell(x_{\text{gene}}, f^{\text{dec}}(f^{\text{enc}}(x_{\text{gene}})))
\end{equation}
where $x_{\text{gene}}$ represents the genomic data, $f^{\text{enc}}$ and $f^{\text{dec}}$ denote the LSTM encoder and decoder. The advantage of this pretraining is that it does not need the labels (e.g., flowering time and grain yield) so we can use the large-scale genomic data to obtain a reasonably good encoder before the training of the predictor. Furthermore, since $x_{\text{gene}}$ is high-dimensional, the gradient vanishing problem may still happen to LSTM even though it is much better than other RNNs. We propose to segment the original high-dimensional input data into several same-size frames, 
\begin{equation}
    x_{\text{gene}} = \left [x^{(0)}_{\text{gene}}, x^{(1)}_{\text{gene}}, x^{(2)}_{\text{gene}}, \ldots, x^{(L)}_{\text{gene}} \right]
\end{equation}
then the encoding of these frames would be trained using the LSTM autoencoder, and the aggregation of frame encodings is finally used as the encoding of the gene as
\begin{equation}
    z_{\text{gene}} = f^{\text{enc}}(x^{(0)}_{\text{gene}}) \oplus f^{\text{enc}}(x^{(1)}_{\text{gene}}) \oplus 
    \cdots \oplus
    f^{\text{enc}}(x^{(L)}_{\text{gene}}).
\end{equation}

\subsection{Genotype-to-phenotype prediction}

With the genomic data encoding from the above component, we propose to learn the unknown relationship between genomic data encoding, additional external variables (i.e., Year, Study, Location, and Season), and phenotypes via a DNN. We use the DNN architecture in this component for two main reasons: 1) as guaranteed by the universal approximation theorem \cite{nishijima2021universal}, a deep neural network with a sufficient number of neurons and the appropriate nonlinear activation function can approximate any continuous function; and 2) there is no explicit knowledge about the targeted relationship, making a DNN a less-biased choice. We tested both multilayer perceptron (MLP) and CNN architectures, and found that the MLP outperformed the CNN (results can be found in the following section).

\section{Experiments}

\subsection{Barley dataset}

We adopted a barley genotype data provided by the Western Crop Genetics Alliance (WCGA) at Murdoch University for modelling the barley genotype-to-phenotype prediction. A total of 894 barley accessions were genotyped using Next-Generation Sequencing. After filtering for heterozygosity, a mapping quality of 20, and a minor allele frequency (MAF) of 0.01, we obtained 30,543 high-quality single nucleotide polymorphism (SNP) markers, which were used as the genotype data. The average density of these genetic markers is approximately 150 kb.

Additionally, we included various environmental variables to account for the growing conditions. These environmental variables encompass location data (five different geographical sites across Western Australia), temporal data (years 2015 and 2016), light conditions (an extended light exposure trial conducted in 2016 at the South Perth site under 18 hours of artificial lighting versus natural light), and agricultural practices (an irrigation trial at Merredin comparing irrigated and non-irrigated conditions). 
The target phenotype variables include `ZS49' and `GrYld(kg/ha)'. 
Table 1 describes the external environmental variables recorded, which are used along with the genotype and phenotype data to build predictive models for barley performance. 

Necessary preprocessing steps were performed based on the data types and specific data problems encountered. The genotype data includes 30,543 SNP markers for distinct barley varieties, with alleles 'A', 'C', 'G', and 'T' encoded numerically, and missing or unavailable alleles assigned a value of -1. Missing data in environmental variables were assigned a value of -1. The resulting processed dataset consists of 4,203 records and 30,554 variables, encompassing genotype, environmental, and phenotype information.

\begin{table}[!t]
    \centering
    \caption{Description of external environmental variables and genotype variables in the adopted dataset}
    \begin{tabular}{l|p{10cm}}
        \hline
        \textbf{Variables} & \textbf{Descriptions} \\ \hline
        Year & The year in which data was recorded, ranging from 2014 to 2016 \\ \hline
        Study & Various conditions to which the crop was exposed, including 18 hrs of artificial lighting, natural lighting, 2 hrs of irrigation, no irrigation, and 1 hr of natural light \\ \hline
        Location & The Western Australia location where the experiments were conducted: Geraldton, Merredin, South Perth, Katanning, and Esperance \\ \hline
        ZS49 & The flowering time of barley, measured in days, a crucial parameter for understanding and predicting crop performance.\\ \hline
        GrYld(kg/ha) & Grain yield measured in kilograms per hectare to quantify the amount of grain produced by a crop per unit area of land.  \\ \hline
    \end{tabular}
    
    \label{tab:variables_description}
\end{table}



\subsection{Experiment settings}

For the LSTM autoencoder, we used two LSTM layers. The lower layer has an input dimension of 1 and a hidden dimension of 128, while the higher layer has an input dimension of 128 and a hidden dimension of 64. The pretraining epoch number is set to 500. 
The default dimension segment is 100, with different values tested and shown in Fig. \ref{fig:ssl}.
The default genomic data embedding dimension is 10, with results for different values shown in Fig. \ref{fig:geneemb}. 
For the MLP, the network width is set to 1,000 with a default depth of 3. Other values are also tested, and the results are shown in Fig.\ref{fig:depth}. 
The activation function used is ReLU. Batch normalization and dropout (with a probability of 0.2) are added between two linear layers. Note that the results in Figs \ref{fig:depth}, \ref{fig:geneemb}, and \ref{fig:ssl} are all from the MLP with the LSTM encoder (without pretraining). The epoch number for model training is 1,000.
In the experiments, the dataset is randomly split into training (80\%) and testing (20\%) sets. The training set is used for model training, and the testing set is used for performance evaluation. We report the average testing results and standard deviations from five runs with five random seeds (123, 124, 125, 126, and 127).

\section{Results}
We evaluated our model's performance (termed as preLSTMMLP) on a processed barley dataset using two prediction tasks: predicting barley flowering time (`ZS49') and grain yield (`GrYld (kg/ha)'). The average Mean Absolute Error (MAE), Root Mean Squared Error (RMSE), and standard deviation results on the testing set are presented in Tables \ref{tab:mae} and \ref{tab:rsme}, respectively.
Our model achieved the lowest MAE and RMSE in both prediction tasks, with an MAE of 7.55 and RMSE of 10.70 for predicting flowering time, and an MAE of 647.36 and RMSE of 843.24 for predicting grain yield. We compared our model against several baseline models, including XGBoost (eXtreme Gradient Boosting)\footnote{https://github.com/dmlc/XGBoost}, which is a widely used ensemble learning method for its high performance and scalability to large datasets \cite{danilevicz2022plant}. XGBoost trained on the full processed dataset resulted in higher MAE and RMSE for both tasks compared to our model.

We also tested an MLP neural network trained directly on the same full processed dataset and without genomic data (termed as MLP(w/o gene)). The results indicated that including genomic data directly in the MLP without the genomic data encoding component led to decreased performance compared to our model in both tasks. Additionally, MLP(w/o gene) performed better than MLP trained on the full processed dataset. This decrease in performance is likely due to the curse of dimensionality, as the genomic data has over $3*10^4$ dimensions.
We also tested our model without pretraining (termed as LSTMMLP) trained on the same full processed dataset. This model performed better than MLP, but not as well as our model with pretraining. Finally, after adding the autoencoder pretraining mechanism, our model's performance significantly improved for both ZS49 and GrYld in terms of MAE and RMSE, demonstrating that the pretraining in the genomic data encoding is crucial as expected.
Additionally, we tested a CNN on the same dataset, with results shown in Table \ref{tab:cnnrsme}. The CNN's performance was inferior to that of the MLP in Table \ref{tab:rsme}. This may be because MLPs tend to perform well on relatively small datasets, whereas CNNs might overfit due to their specialized architecture and thus require more data to learn meaningful features.

\begin{table}[h]
\caption{Predictive results (MAE) using MLP in predicting flowering time ('\textit{ZS49}') and grain yield ('\textit{GrYld}')}
\centering
\renewcommand{\arraystretch}{2} 
\resizebox{\textwidth}{!}{%
\begin{tabular}{c  c c c c c }
\hline
\textbf{Outputs} &  \textbf{XGBoost} & \textbf{MLP(w/o gene)} & \textbf{MLP} & \textbf{LSTMMLP} & \textbf{preLSTMMLP} \\
\hline\hline
\textit{ZS49} &  16.30 $\pm$ 0.5 &  8.10 $\pm$ 0.1 &  11.57 $\pm$ 0.2 & 8.05 $\pm$ 0.2  &  \textbf{7.55 $\pm$ 0.3}  \\
\textit{GrYld} &  727.98 $\pm$ 20.7 &  694.60 $\pm$ 6.9 &  733.74 $\pm$ 18.4 &  710.61 $\pm$ 24.9 &  \textbf{647.36 $\pm$ 8.0}  \\
\hline
\end{tabular}%
}
\label{tab:mae}
\end{table}

\begin{table}[h]
\caption{Average predictive results (RMSE) using MLP in predicting flowering time ('\textit{ZS49}') and grain yield ('\textit{GrYld}')}
\centering
\renewcommand{\arraystretch}{2} 
\resizebox{\textwidth}{!}{%
\begin{tabular}{c  c c c c c }
\hline
\textbf{Outputs} &  \textbf{XGBoost} & \textbf{MLP(w/o gene)} & \textbf{MLP} & \textbf{LSTMMLP} & \textbf{preLSTMMLP} \\
\hline
\hline
\textit{ZS49} &  23.61 $\pm$ 0.9 &  11.32 $\pm$ 0.3 &  15.13 $\pm$ 0.3 &  11.30 $\pm$ 0.3 &  \textbf{10.70 $\pm$ 0.4}  \\
\textit{GrYld} &  954.84 $\pm$ 30.7 & 904.02 $\pm$ 14.5 &  941.79 $\pm$ 30.9 &  911.12 $\pm$ 45.5 &  \textbf{843.24 $\pm$ 8.9}  \\
\hline
\end{tabular}
}
\label{tab:rsme}
\end{table}

\begin{table}[h]
\caption{Average predictive results (RMSE) using CNN in predicting flowering time ('\textit{ZS49}') and grain yield ('\textit{GrYld}')}
\centering
\renewcommand{\arraystretch}{2} 
\begin{tabular}{c c c c c }
\hline
\textbf{Outputs}~~  & \textbf{CNN(w/o gene)}~ & \textbf{CNN}~ & \textbf{LSTMCNN}~ & \textbf{preLSTMCNN} \\
\hline\hline
\textit{ZS49} &   17.73 $\pm$ 0.8 &  18.98 $\pm$ 2.9 &  16.56 $\pm$ 0.7 &  17.13 $\pm$ 0.9  \\
\textit{GrYld} &   974.50 $\pm$ 14.0 &  1071.81 $\pm$ 57.0 &  955.39 $\pm$ 12.9 &  1020.15 $\pm$ 106.3  \\
\hline
\end{tabular}
\label{tab:cnnrsme}
\end{table}

\begin{figure}[h]
\centering
\includegraphics[scale=0.5]{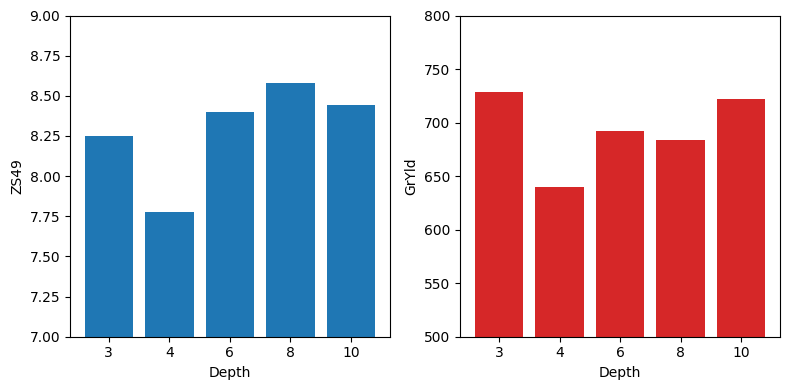}
\caption{Impact of MLP Depth on predictive results (MAE): Left - ZS49, Right - GrYld}
\label{fig:depth}
\end{figure}

\begin{figure}[h]
\centering
\includegraphics[scale=0.5]{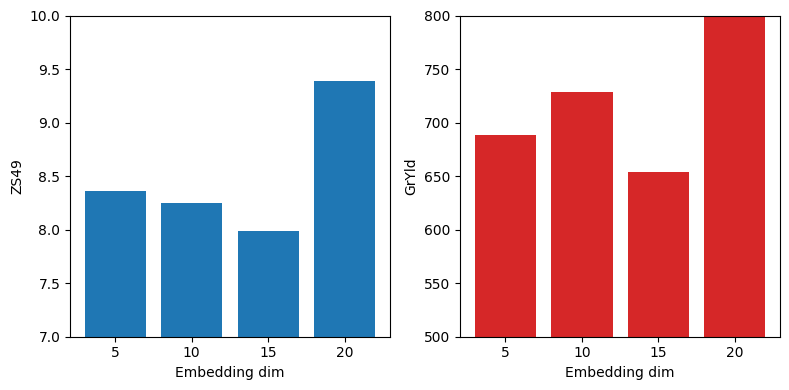}
\caption{The impact of gene embedding dimension on predictive results (MAE): Left - ZS49, Right - GrYld}
\label{fig:geneemb}
\end{figure}

\begin{figure}[h]
\centering
\includegraphics[scale=0.5]{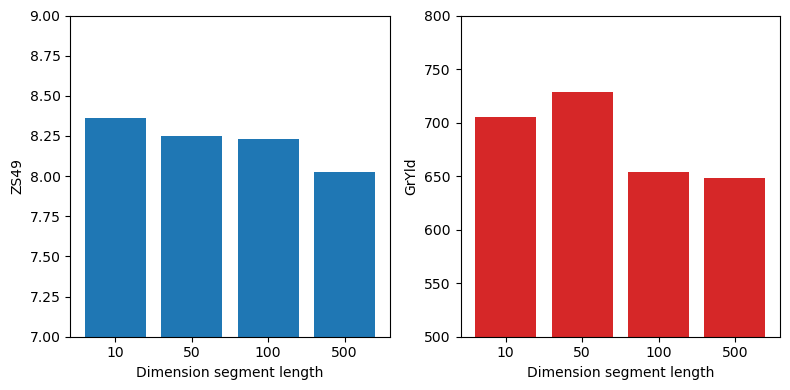}
\caption{The impact of dimenision segment length on predictive results (MAE): Left - ZS49, Right - GrYld}
\label{fig:ssl}
\end{figure}

We further illustrate the impact of parameters including MLP depth, gene embedding dimension, and dimension segment length on prediction performance in Figs. \ref{fig:depth}, \ref{fig:geneemb} and \ref{fig:ssl}, respectively. Our findings indicate that an MLP depth of 4 achieved the lowest MAE for both prediction tasks, striking a balance between model complexity and predictive performance.
For the gene embedding dimension, a value of 15 yielded the lowest MAE for both outcomes. However, increasing the dimension to 20 significantly increased the MAE for both tasks, while reducing it to 10 also led to higher MAE values. This suggests a sensitive selection is necessary for this parameter to optimize model performance.
In terms of dimension segment length, a length of 500 resulted in the lowest MAE for both outcomes, compared to shorter lengths ranging from 10 to 100. This indicates that the LSTM encoder architecture more effectively captures relationships within the higher dimension data.

\section{Conclusion}
We proposed a new LSTM autoencoder-based DNN model for crop genotype-to-phenotype prediction, and applied it to predicting barley genotype to flowering time and grain yield with improved performance. Specifically, we introduced genomic data encoding by pretraining the LSTM using an autoencoder structure before phenotype prediction to extract latent feature representations from the complex high-dimensional genomic data. Our model achieved the lowest MAE and RMSE in both prediction tasks compared to other baseline models, demonstrating its potential to enhance predictive power in handling complex datasets encompassing genotype, phenotype, and environmental data in the agriculture context. In the future, we plan to include time series environmental variables such as soil temperature and rainfall based on locations to further enhance the model's prediction performance. We will also expand the comprehensive testing of our models on different crop types.

%
%
%
%

\bibliographystyle{IEEEtran} 
\bibliography{paper} 

\end{document}